\documentclass[10pt]{article}
\usepackage[margin=1.0in]{geometry}
\usepackage{epsfig,ulem,amssymb,soul}
\usepackage{hyperref}
\usepackage{graphicx}
\usepackage[skip=2pt]{caption}
\usepackage{epstopdf}
\usepackage{amsfonts,amsmath}
\usepackage{bm}
\usepackage{setspace}
\usepackage{comment}
\usepackage{siunitx}
\usepackage{mathrsfs}
\usepackage{subcaption}
\usepackage[hang,flushmargin,bottom]{footmisc}
\usepackage{mathtools}
\usepackage{lipsum}
\usepackage{array}
\usepackage{enumerate}
\usepackage{ifthen}
\usepackage[affil-it]{authblk}
\usepackage{cite}
\usepackage{leftidx}
\usepackage{relsize}
\usepackage{braket}
\usepackage[ruled,vlined]{algorithm2e}
\usepackage{float}
\usepackage[right]{lineno}

\usepackage[autostyle]{csquotes}
\usepackage[dvipsnames]{xcolor}
\providecommand{\keywords}[1]{\textbf{\textit{Keywords-}} #1}

\title{Variable-Order Fracture Mechanics \\and its Application to Dynamic Fracture}
\author[1]{Sansit Patnaik}
\author[1]{Fabio Semperlotti}
\affil[1]{School of Mechanical Engineering, Ray W. Herrick Laboratories, Purdue University, West Lafayette, IN 47907}
\date{}

\begin{document}
\maketitle
\begin{abstract}

This study presents the formulation, the numerical solution, and the validation of a theoretical framework based on the concept of variable-order mechanics and capable of modeling dynamic fracture in brittle and quasi-brittle solids. More specifically, the reformulation of the elastodynamic problem via variable and fractional order operators enables a unique and extremely powerful approach to model nucleation and propagation of cracks in solids under dynamic loading. The resulting dynamic fracture formulation is fully evolutionary hence enabling the analysis of complex crack patterns without requiring any \textit{a priori} assumptions on the damage location and the growth path, as well as the use of any algorithm to track the evolving crack surface. The evolutionary nature of the variable-order formalism also prevents the need for additional partial differential equations to predict the damage field, hence suggesting a conspicuous reduction in the computational cost. Remarkably, the variable order formulation is naturally capable of capturing extremely detailed features characteristic of dynamic crack propagation such as crack surface roughening, single and multiple branching. The accuracy and robustness of the proposed variable-order formulation is validated by comparing the results of direct numerical simulations with experimental data of typical benchmark problems available in the literature.\\

\noindent\keywords{Fractional calculus, Variable-order operators, Fracture mechanics, Dynamic fracture}\\
\noindent$^\dagger$ All correspondence should be addressed to: \textit{spatnai@purdue.edu} or \textit{fsemperl@purdue.edu}\\

\end{abstract}

\section*{Introduction}
Fracture is one of the most commonly encountered mode of failure in structural systems across a broad spectrum of applications spanning the civil, mechanical, and aerospace engineering fields. The prevention of fracture-induced failure is a major concern of structural design and has historically motivated the development of theoretical and experimental methodologies to predict nucleation and propagation of structural damage. While the general topic of fracture mechanics is very complex in itself due to the coexistence of multiple physical processes occurring over multiple spatial scales, the specific topic of dynamic fracture is possibly even more challenging due to the occurrence of crack surface roughening, instabilities, and branching. Detailed discussions on the implications and modeling approaches for dynamic fracture can be found in many sources such as \cite{ravi2004dynamic}. During the last few decades, the analysis of dynamic fracture has certainly largely benefited and made significant progress thanks to the rapid development of numerical methods.
From a high level perspective, the approaches available for the analysis of damage can be divided into two categories, namely, discrete and continuum. This classification refers specifically to the modeling of the damage interface so that, while in both cases the solid is treated as a continuum, in the former class of approaches the displacement is modeled as a discontinuous field across the fracture surface. In the latter category, instead, the displacement is treated as continuous field everywhere (even across the crack surface), but the local value of the elastic energy is reduced by accounting for the softening of the material properties associated with the fracture-induced degradation. In the following, we briefly review some of the most accredited dynamic fracture models in order to clearly define the context in which our variable-order approach is defined.

Discrete approaches to the modeling of dynamic fracture include extended finite element methods (XFEM) \cite{belytschko2003dynamic,remmers2003cohesive,rabczuk2004cracking,song2008comparative}, discontinuous cell methods (DCM) \cite{cusatis2017discontinuous}, cohesive interface element techniques \cite{xu1994numerical,park2012adaptive,spring2018achieving}, discontinous Galerkin methods \cite{abedi2010adaptive,abedi2017effect}, and lattice based models \cite{wang2009hybrid,chakraborty2013pseudo}. From a general perspective, these approaches are based either on linear elastic fracture mechanics (LEFM) \cite{griffith1921vi} or on the cohesive zone model (CZM) \cite{barenblatt1959formation}. Owing to its computational and multiscale analysis capabilities, the XFEM has quickly risen in popularity and it is currently one of the most widely used approaches. In XFEM, cracks are represented as discrete discontinuities that are embedded in the damaged elements by enriching the displacement field according to the method of partition of unity \cite{babuvska1997partition}. This approach implies that the front of the discontinuity (i.e. the crack) must be tracked explicitly. While several tracking algorithms have been proposed over time \cite{belytschko2000element,rabczuk2004cracking,song2008comparative}, the front tracking process is quite computationally intensive, particularly in three-dimensional problems involving complex crack topology. Another important limitation consists in the need for a branching criterion which is often \textit{ad-hoc} and limited to two crack branches. Exceptions to this latter comment are the formulations based on either the DCM \cite{cusatis2017discontinuous} or the interface elements \cite{xu1994numerical,nguyen2014discontinuous}, which however must be inserted in the model \textit{a priori} hence posing the problem of knowing the location and path of propagation of damage. The front tracking limitation was addressed by the use of lattice models where the continuum is replaced by a system of rigid particles that interact via a network of linear and nonlinear springs. More recently, Silling \cite{silling2000reformulation,silling2005meshfree} proposed an approach denominated \textit{peridynamics} that models the solid medium as a nonlocal lattice of particles described via an integral formulation. During the last two decades this approach has received much attention and it has been used in many diverse applications. In the context of dynamic fracture, peridynamics has shown to be able to address several of the above shortcomings and could accurately capture crack intersections and branching in complex structures and materials. An important caveat of the lattice models derives from the fact that the springs stiffness is often defined heuristically and various elastic phenomena (e.g. the Poisson’s effect) cannot be reproduced exactly.

In the second category, the continuum approaches, we find the crack band method \cite{bavzant1983crack}, nonlocal integral damage models \cite{pijaudier1987nonlocal}, nonlocal stress based damage models \cite{pereira2017numerical}, and the more recent class of phase-field models \cite{francfort1998revisiting,aranson2000continuum,miehe2010phase,bourdin2011time,hofacker2012continuum,wu2017unified,nguyen2018modeling}. Phase-field models are undoubtedly the methods that have seen the highest popularity given their overall accuracy and ease of implementation. In phase-field models, sharp cracks are regularized by a diffused damage field while a variational approach is adopted to obtain evolutionary equations for both the displacement and the damage fields \cite{francfort1998revisiting}. The formulation also includes a small and positive length scale parameter so that, in the limit for the parameter approaching zero, the phase-field representation of the crack converges to the original problem of a sharp crack \cite{ambrosio1990approximation}. The use of a phase-field regularization prevents the need for an explicit tracking of the crack surface discontinuity. It follows that the numerical implementation of the phase-field model is relatively straight-forward when compared to the previously mentioned discrete approaches. An important disadvantage of these models lies in their high computational cost, which follows from the need to solve a coupled system of partial differential equations for both the damage (phase) and the displacement fields \cite{nguyen2018modeling}. This limitation becomes even more significant when the phase field approach is applied for fracture analysis in three-dimensional media. Additionally, phase-field models are subject to an \textit{artificial} widening effect in the damaged area at the point of occurrence of instability \cite{borden2012phase,nguyen2018modeling}, which is in contrast to the microbranching and crack surface roughening seen in experiments \cite{ravi1984experimental}. A detailed review on phase-field models can be found in \cite{wu2018phase}.

In very recent years, variable-order fractional calculus (VO-FC) has emerged as a powerful mathematical tool to model a variety of discontinuities and nonlinear phenomena. Variable-order (VO) fractional operators are a natural extension of constant-order fractional operators that allow the differentiation and integration of functions to any real or complex valued order \cite{podlubny1998fractional}. In VO operators, the order can be function of time, space, internal variables (e.g. system energy or stress) or even of external variables (e.g. temperature or external mechanical loads) \cite{lorenzo2002variable}. As the VO-FC formalism allows updating the system's order depending on either its instantaneous or historical response, the corresponding model can evolve seamlessly to describe widely dissimilar dynamics without the need to modify the structure of the underlying governing equation. Thus, a very significant feature of VO-based physical models consists in their evolutionary nature; a property that can play a critical role in the simulation of nonlinear dynamical systems \cite{lorenzo2002variable,coimbra2003mechanics,patnaik2020application}. In recent years, many applications of VO-FC to practical real-world problems have been explored including, but not limited to, modeling of anomalous diffusion in complex structures with spatially and temporally varying properties \cite{Chechkin1}, the response of nonlinear oscillators with spatially-varying constitutive law for damping \cite{coimbra2003mechanics}, nonlinear dynamics with contacts and hysteresis \cite{patnaik2020application}, and even complex control systems \cite{Ostalczyk1}. The interested reader can find a comprehensive review of the applications of VO-FC in \cite{patnaik2020review}.

In this study, we present a theoretical and computational framework based on VO fractional operators and capable of effectively capturing the many features of dynamic fracture in brittle and quasi-brittle solids. We will show how the many unique capabilities of this framework build directly on the several remarkable properties afforded by these fascinating mathematical operators. Dynamic fracture is a quintessential evolutionary nonlinear dynamic problem that involves the propagation of nonlinearities and discontinuities through a system. VO fractional operators are uniquely equipped to model these complex class of dynamical problems. The VO framework presented in this paper builds upon the mathematical structure presented in \cite{patnaik2020variable} which focused on the modeling of the propagation of dislocations through lattices of particles using physics-informed order variations. More specifically, the VO model introduced in \cite{patnaik2020variable} leveraged an order variation law based on the relative displacements of particles within the lattice in order to capture the formation and annihilation of pair-wise bonds. The general strategy followed that outlined in \cite{patnaik2020application,patnaik2020modeling} for physics-driven VO laws for discrete systems. The approach resulted in the formulation of evolutionary VO fractional differential equations capable of capturing the transition towards a nonlinear dynamic regime (associated with the motion of dislocations) without having to explicitly track the location of the dislocation. In this study, we extend this general approach to continuous systems by formulating a VO elastodynamic framework uniquely suited for the analysis of dynamic fracture and capable of detecting the formation and propagation of damage by means of a strain-driven order variation laws. The introduction of VO operators in the continuum elastodynamic formulation allows the governing equations to evolve (from linear to nonlinear) and adapt (by capturing discontinuities) based on both the local response and the underlying damage mechanism while eliminating the need for explicitly tracking the damage front. 
We will show that the resulting formulation is capable of capturing key features associated with the dynamic fracture mechanism such as roughening of the crack surface, crack instability, and crack branching without the need of any \textit{a-priori} assumptions or \textit{ad-hoc} criteria. Further, contrary to phase-field models, no additional partial differential equations are needed to predict the evolution of the damage field. Indeed, in the VO framework, the damage field evolves naturally guided by the variation of the order of the fractional operators that solely depends on the instantaneous response of the system. In the second part of this study, the VO dynamic fracture model is validated by applying it to the direct numerical simulation of three benchmark experiments available in the literature: 1) the Kalthoff-Winkler experiment that involves the impact shear loading of a doubly notched specimen \cite{kalthoff1988failure}, 2) the dynamic crack branching experiment \cite{ravi1984experimental}, and 3) the John-Shah experiment that involves the impact loading of a pre-notched concrete slab \cite{john1990mixed}.

\section*{Material and Methods}

We briefly discuss the general strategy leading to the formulation of evolutionary governing equations based on VO Riemann-Liouville (VO-RL) derivatives of constants \cite{patnaik2020application,patnaik2020modeling}. Then, we apply these operators to formulate an evolutionary elastodynamic framework suitable for the modeling of dynamic fracture. Some background and discussion of the fundamental properties of the VO-RL operators used in this study are provided in Supplementary Information (SI).\\

\noindent\textbf{Evolutionary governing equations via VO-RL derivatives of constants:} A particularly interesting property of fractional-order Riemann-Liouville operators stems out of their behavior when applied to the fixed-order derivative of a constant. It is found that this fractional order derivative is not equal to zero, unless the order converges to an integer. While this is an unexpected and maybe even unsettling property of such operators, at least in view of classical integer order calculus, we will show that this property has extremely valuable implications for modeling physical systems exhibiting highly nonlinear and discontinues behavior. Mathematically, the RL derivative of a constant $A_0$ $\in \mathbb{R}$ to a constant fractional-order $\alpha_0$ $\in \mathbb{R^+}$ defined on the interval $(a,t) \in \mathbb{R}$ is given as \cite{podlubny1998fractional}:
\begin{equation}
\label{eq: CORL}
D^{\alpha_0}_tA_0=\frac{A_0~(t-a)^{-\alpha_0}}{\Gamma(1-\alpha_0)}
\end{equation}
where $\Gamma(\cdot)$ is the Gamma function. Note that, although apparently non intuitive, this is merely an intrinsic property of the RL operator. The use of this property was originally outlined and extended to variable-order in \cite{patnaik2020application,patnaik2020modeling} where it was applied to the modeling of highly nonlinear mechanisms in dynamical systems. More specifically, the properties offered by the variable-order Riemann-Liouville (VO-RL) derivative of a constant creates a unique opportunity to formulate  governing equations in an evolutionary form. In the following, we briefly review these characteristics in order to lay the necessary foundation for the development of the VO elastodynamic formulation. 

Consider a function $\alpha(t)$ constructed using a continuous real-valued function $\kappa(t)$ in the following fashion \cite{patnaik2020application}:
\begin{equation}
\label{eq: VO_definition}
\alpha(t)=\exp({-{\kappa_0}\kappa(t)})
\end{equation}
where the function $\kappa(t)$ is some function designed to capture the desired physical mechanism of interest and the one producing the order variation. Specific details on the selection of this function in the context of fracture mechanics will be provided when addressing the VO dynamic fracture formulation. We emphasize that, while the characteristic function $\kappa(t)$ introduced above was defined to be a function of time $t$, the functional dependence can be extended to include any other dependent or independent variables depending on the specific physical problem. Further, $\kappa_0 \in \mathbb{R}^+$ is a scaling factor that allows calibrating the order variation on the scale of the characteristic response of the physical system. A detailed discussion of the procedure to determine the value of $\kappa_0$ is outlined in the SI along with an illustrative example. For a given $\kappa_0$, the limiting behavior of $\alpha(t)$ is:
\begin{equation}
\label{eq: VO_definition_step1}
\alpha(t) \rightarrow \left\{ \begin{matrix} \infty & \kappa(t) < 0\\
0 & \kappa(t)>0\end{matrix} \right.
\end{equation}
Now, we can indicate the VO-RL derivative of the constant $A_0$ to the order $\alpha(t)$ on the interval $(a,t)$ as $_a D^{\alpha(t)}_t f(t)$ or, in the interest of a more compact notation, as $D^{\alpha(t)}_t f(t)$. Equations~(\ref{eq: CORL}-\ref{eq: VO_definition_step1}) lead to:
\begin{equation}
\label{eq: VO_definition_step2}
_a D^{\alpha(t)}_t f(t) \equiv D^{\alpha(t)}_t A_0=
\left\{\rule{0cm}{0.9cm}\right. \begin{matrix} \lim\limits_{\alpha(t)\rightarrow\infty} \frac{A_0~(t-a)^{-\alpha(t)}}{\Gamma[1-\alpha(t)]} = 0 & \kappa(t)\leq0\\
\lim\limits_{\alpha(t)\rightarrow 0} \frac{A_0~(t-a)^{-\alpha(t)}}{\Gamma[1-\alpha(t)]} = A_0 & \kappa(t)>0 \end{matrix} 
\end{equation}
It appears that, when the VO-RL operator is applied to a constant under the conditions in Eq.~(\ref{eq: VO_definition}), a discontinuous (switch-like) behavior can be captured simply following a change in sign of the function $\kappa(t)$. It is exactly this switching behavior that can be exploited to simulate the occurrence of certain nonlinear and discontinuous dynamical properties of mechanical systems. More specifically, consider defining VO operators as part of a governing equation such that its variation can capture changes in the properties of the systems such as, for example, a change in stiffness (e.g. bilinear stiffness) or the occurrence of geometric discontinuities (e.g. dislocations in a lattice or crack in a continuum). In all these cases, the response of the system changes from initially linear to, potentially, highly nonlinear. The onset of either types of nonlinearities or discontinuities results in an implicit reformulation of the underlying system dynamics. This change in the underlying dynamics can be captured in the order $\alpha(t)$ via the function $\kappa(t)$. It immediately follows that a change in the order $\alpha(t)$ results in an implicit reformulation of the equations of motion following a change in the underlying physical mechanisms dominating the response of the system. This characteristic was illustrated to formulate evolutionary equations to model contact dynamics, hysteretic behavior \cite{patnaik2020application}, and motion of edge-dislocations in lattice structures \cite{patnaik2020variable}. In the present study, we extend this unique behavior of the VO-RL operator to simulate the initiation and propagation of cracks in solids. Such behavior is achieved by proper integration of the VO-RL operators in the elastodynamic formulation.\\ 

\noindent\textbf{VO elastodynamic formulation:} The strong form of the governing equation for a solid having a volume $\Omega$ (see Fig.~(S3) in SI) is given in the well-known form:
\begin{equation}
    \label{eq: elastodynamic_eqs}
    \nabla\cdot\bm{\sigma} + \bm{f}=\rho\bm{\ddot{u}}
\end{equation}
where $\bm{\sigma}$ denotes the stress field, $\bm{u}$ denotes the displacement field, $\bm{f}$ denotes the externally applied force, and $\rho$ denotes the density of the solid. The bold-face is used to indicate either vectors or tensors. The above equations of motion are subject to the following boundary (BC) and initial (IC) conditions:
\begin{subequations}
\label{eq: BC_IC}
\begin{equation}
    \text{BC}: \left \{ \begin{matrix}
    \bm{u}(\bm{x},t)=\bm{\overline{u}}(\bm{x},t) &  \bm{x}\in\partial\Omega_u \\
    \bm{\hat{n}}\cdot\bm{\sigma}=\bm{\overline{t}}(\bm{x},t) & \bm{x}\in\partial\Omega_t 
    \end{matrix}
    \right.
\end{equation}
\begin{equation}
\text{IC}: \left \{ \begin{matrix}
    \bm{u}(\bm{x},0)=\bm{u}_0 (\bm{x}) & \bm{x}\in\Omega \\
    \bm{\dot{u}}(\bm{x},0)=\bm{v}_0 (\bm{x}) & \bm{x}\in\Omega 
    \end{matrix}
    \right.
\end{equation}
\end{subequations}
where $\partial\Omega_u$ and $\partial\Omega_t$ denote the portion of the boundary where essential and natural boundary conditions are applied, respectively. $\bm{\overline{u}}$ and $\bm{\overline{t}}$ denote the externally applied displacement and traction at $\partial\Omega_u$ and $\partial\Omega_t$, respectively. While $\bm{u}_0$ and $\bm{v}_0$ denote the displacement and velocity fields at $t=0$. The stress developed in the medium upon damage is defined in the following fashion:
\begin{equation}
    \label{eq: stress_damage_relation}
    \bm{\sigma} = \psi(d) \bm{\mathbb{C}}:\bm{\varepsilon}
\end{equation}
where $\bm{\mathbb{C}}$ is the classical fourth-order elasticity tensor and $\bm{\varepsilon}$ is the symmetric displacement-gradient strain tensor. $\psi(d)$ is a degradation function of the damage variable $d\in[0,1]$ such that $\partial\psi/\partial d\leq 0$; this latter condition originates from the thermodynamic consideration that the degradation function must lead to a decrease in the elastic energy with an increase in damage size.
In this study, the damage variable $d$ is defined such that $d=0$ indicates the undamaged state, while $d=1$ indicates a fully damaged state. We note that the same stress-damage-strain constitutive relation has also been adopted in several classical dynamic fracture formulations \cite{wu2017unified,cusatis2017discontinuous,nguyen2018modeling}.

In the VO dynamic fracture formulation, we adopt a strain-based criterion to detect the onset of damage. More specifically, damage at a given point occurs when the maximum principal strain at the given point exceeds a critical strain derived from the elastic strength of the material. The VO-RL formalism presented previously allows us to define the characteristic function $\kappa(\bm{x},t)$ which allows detecting the onset of damage following the strain-driven physical law. More specifically, we define a VO $\alpha(\bm{x},t)$ in the following manner:
\begin{equation}
    \label{eq: VO_damage_def}
    \alpha(\bm{x},t) = \exp \bigg[ - \kappa_0 \underbrace{\left(  \frac{ \overline{\varepsilon}(\bm{x},t) - \varepsilon_u  } {\varepsilon_u} \right)}_{\kappa(\bm{x},t)} \bigg]
\end{equation}
where $\kappa_0$ is the previously introduced scaling factor.
$\varepsilon_u \in \mathbb{R}^+$ is the material parameter defining the ultimate tensile strain limit governing the onset of damage. $\overline{\varepsilon}(\bm{x},t)$ is the maximum principal strain that occurs at a given point $\bm{x}$ until the instant $t$. More specifically:
\begin{equation}
    \label{eq: history}
    \overline{\varepsilon}(\bm{x},t) = \max_{\tau\in[0,t]} \tilde{\varepsilon}(\bm{x},\tau) = \max_{\tau\in[0,t]} \left[ \max \left\{ \tilde{\varepsilon}_x(\bm{x},\tau), \tilde{\varepsilon}_y(\bm{x},\tau), \tilde{\varepsilon}_z(\bm{x},\tau) \right\} \right]
\end{equation}
where $\tilde{\varepsilon}(\bm{x},\tau)$ denotes the maximum principal strain component (i.e., maximum of different eigen strain values $\tilde{\varepsilon}_x(\bm{x},\tau), \tilde{\varepsilon}_y(\bm{x},\tau), \tilde{\varepsilon}_z(\bm{x},\tau)$) at the point $\bm{x}$ and at a given time instant $\tau$. 
Recall that a change in the sign of the argument $\kappa(\bm{x},t)$ within the exponential of the VO results in a reformulation of the underlying governing equations. Exploiting the previously described property of VO-RL operators and defining a physics-driven variation of the order according to Eq.~(\ref{eq: VO_damage_def}), the damage variable can be written as:
\begin{equation}
    \label{eq: damage_definition}
    d (\bm{x},t) = \underbrace{D^{\alpha(\bm{x},t)}_t d_0}_{\text{Term I}} - \underbrace{D^{\alpha(\bm{x},t)}_t \varepsilon_u \left[ \frac{1}{ \overline{\varepsilon}(\bm{x},t) } \exp \left( - \frac{(\overline{\varepsilon}(\bm{x},t) - \varepsilon_u )} {\varepsilon_R} \right)  \right]}_{\text{Term II}}
\end{equation}
where $d_0=1$ indicates the maximum possible damage. Before discussing the specific role of the two terms in Eq.~(\ref{eq: damage_definition}), we explain the different parameters introduced in the equation. $\varepsilon_R$ is defined as:
\begin{equation}
    \label{eq: rate}
    \varepsilon_R = 2\varepsilon_u \left( 1 - \frac{l_f}{l_t} \right)
\end{equation}
where $l_t = 2EG_f/\sigma_u^2$ is the characteristic material length for an isotropic solid having Young's modulus $E$, fracture energy $G_f$, and elastic strength $\sigma_u$ \cite{baz1984rock}. 
$l_f$ determines a characteristic physical dimension of the area within which the crack is localized and, in numerical implementations, it is directly related to the size of the elements used for the spatial discretization of the domain. In other terms, $l_f$ dictates the width of the crack path at a given point, that is the distance perpendicular to the crack path at the same point, within which the damage varies between its extreme values.
Further, the parameter $\varepsilon_R$ governs the damage evolution rate that determines the level of damage via term II (see also, Fig.~(S3) of SI). In order to guarantee the insensitivity of the results to the specific choice of the numerical mesh adopted, it is necessary that $l_f<l_t$ \cite{baz1984rock,cusatis2017discontinuous}. The latter condition also follows from the fact that the size of the elements used to simulate the crack must be smaller than the characteristic material length for accurate resolution of the crack path.

It follows from Eqs.~(\ref{eq: VO_damage_def},\ref{eq: damage_definition}) that $d(\bm{x},t) = 0$ for $\overline{\varepsilon} (\bm{x},t) \leq \varepsilon_u$ and $d(\bm{x},t) \rightarrow 1$ when $\overline{\varepsilon} (\bm{x},t) \gg \varepsilon_u$.
Thus, it appears that, when the maximum principal strain $\overline{\varepsilon}(\bm{x},t)$ exceeds the critical strain limit $\varepsilon_u$, damage is initiated at that particular point. The specific value of the damage variable is determined by the combined effects of the two terms in Eq.~(\ref{eq: damage_definition}). While term I in Eq.~(\ref{eq: damage_definition}) sets the value of the maximum damage, term II allows for an exponential interpolation of the damage between its extreme values (0 and 1) depending on the amount by which the maximum principal strain ($\overline{\varepsilon}$) exceeds the critical strain ($\varepsilon_u$) by using the parameter $\epsilon_R$. Note that the evolution of both these terms is guided by the VO-RL derivatives.
More specifically, the VO-RL operator allows detecting the onset of damage in the solid driven by the VO $\alpha(\bm{x},t)$. This leads to an automatic reformulation of the underlying governing equations via Eqs.~(\ref{eq: elastodynamic_eqs},\ref{eq: damage_definition}) in order to account for the occurrence and evolution of damage. Remarkably, the resulting evolutionary VO model does not require any front tracking algorithm nor additional criteria to capture the characteristic features of the dynamic crack mechanism such as roughening and branching. This latter comment will be more evident when contrasting the above formulation with the numerical results presented below. 

The VO dynamic fracture formalism deserves some additional remarks. First, note that the constitutive relations defined in Eq.~(\ref{eq: stress_damage_relation}) result in an identical tensile and compressive fracture behaviour which is not generally true when modeling the failure of brittle and quasi-brittle solids. Several researchers have captured the asymmetric tensile/compressive damage by performing a spectral decomposition of the strain energy density and by degrading only the positive strain energy \cite{miehe2010phase,borden2012phase}. In this study, we incorporate this asymmetric behavior via the maximum principal strain based damage criterion, which follows from the well known Rankine criterion \cite{nguyen2018modeling}. In other terms, as described previously, the crack is allowed to nucleate only when the maximum principal strain exceeds the critical tensile strain ($\varepsilon_u$). This specific feature ensured via the VO defined in Eq.~(\ref{eq: VO_damage_def}) allows to model the asymmetric damage behaviour.
Further, the definition of the parameter $\overline{\varepsilon}$ in Eq.~(\ref{eq: history}) based on its past history, along with the condition $\dot{d}(\bm{x},t) \geq 0$, ensure irreversibility of the system. More specifically, these conditions ensure that the length of the crack, denoted by $\Gamma_c$, is monotonically increasing, that is $\Gamma_c(t_1) \subseteq \Gamma_c(t_2)~\forall~ t_1<t_2$. 
Additionally, the use of the strain-history based parameter $\overline{\varepsilon}$ leads to simpler numerical implementations as it allows for an operator split algorithm within a given time-step, wherein the displacement field and the damage field are updated in a staggered manner. The same concept, albeit using a strain-energy based history variable, is often employed in phase-field models of dynamic fracture \cite{miehe2010phase,hofacker2012continuum}. Following this staggered numerical implementation,  the computation of the damage field is a purely algebraic operation and it does not require the minimization of an additional potential function which, in the case of phase-field models, corresponds to the crack surface density function. The most immediate consequence is that the VO approach reduces the computational cost of dynamic fracture when compared to phase-field models.

Finally, note that the damage value $d=0$ obtained for $\overline{\varepsilon}=\varepsilon_u$ indicates a zero crack-tip opening displacement. As $\overline{\varepsilon}$ increases, the damage increases leading to an increase in the crack-tip displacement. Since the crack-tip opening displacement is directly linked to the strain value, it follows that the ratio of the instantaneous crack-tip opening displacement to the maximum crack-tip opening displacement is equal to the ratio of the instantaneous strain to the maximum strain (obtained for $d=1$). Further, given the strain-based definition of the damage variable in Eq.~(\ref{eq: damage_definition}), it follows that the ratio of the crack-tip displacement $(\delta_c)$ to the critical crack-tip displacement $(\delta_0)$ can be given as the ratio of the damage parameters $d/d_0$. This formulation allows expressing different traction separation laws (also called softening laws) directly in terms of the damage variable $d$ without the need for additional characteristic functions similar to \cite{cusatis2017discontinuous}. In this study, we use two different traction separation laws expressed in terms of the damage variable as:
\begin{subequations}
\label{eq: softening_laws}
\begin{equation}
    \label{eq: softening_law_linear}
    \psi(d) = 1-d
\end{equation}
\begin{equation}
    \label{eq: softening_law_cornelissen}
    \psi(d) = (1 + \eta_1^3 d^3) \exp(-\eta_2 d) - d(1+\eta_1^3) \exp(-\eta_2)
\end{equation}
\end{subequations}
Equation~(\ref{eq: softening_law_linear}) is a linear law for brittle materials \cite{wu2017unified}, while Eq.~(\ref{eq: softening_law_cornelissen}) is the Cornelissen law \cite{cornelissen1986experimental} for concrete (a quasi-brittle material). This latter law was obtained experimentally in \cite{cornelissen1986experimental} and the coefficients were found to be $\eta_1=3$ and $\eta_2=6.93$. Note that, in both cases, $\psi(d)$ is bounded so that $\psi(d)\in[0,1]$ for $d\in[0,1]$.

We highlight that the VO elastodynamic formulation described above does not account for contact conditions, such as those that occur when the free surfaces of a crack come in contact when subjected to compressive loads. Note that this is not a limitation of the methodology but merely a decision of the authors to focus this work on aspects concerning crack initiation and propagation. Indeed, the contact problem is typically not addressed in classical treatments of dynamic fracture. However, the VO formulation can easily account for contact dynamics by simply adding dedicated terms in the VO derivative. The case of contact via VO operators was previously treated by the authors in \cite{patnaik2020modeling}, albeit only for discrete systems.


\section*{Results}
To demonstrate the accuracy and the robustness of the VO fracture mechanics framework, we apply it to perform numerical simulations of three classical benchmark experiments available in the literature. The first two examples refer to the Kalthoff-Winkler experiment \cite{kalthoff1988failure} and the classical crack-branching experiment \cite{ravi1984experimental}; both pertain to the dynamic fracture of brittle solids. The last example consists of mixed-mode fracture of a concrete slab under an impact load that was analyzed experimentally in \cite{john1990mixed}. In all three examples, the numerical results were obtained using a plane strain elastic model. The computational domain was discretized using uniform quadrilateral elements and the dynamic solution was computed using an explicit Newmark solver. Further, a lumped mass matrix was used in the dynamic solver to suppress high-frequency elastic oscillations (noises) and to ensure conservation of energy. The time-step ($\Delta t$) used in the dynamic solver, was determined using the Courant-Friedrichs-Lewy (CFL) condition. For a more conservative scheme, we used $\Delta t = 0.9 \Delta t_0 = 0.9h/c$, where $h$ denotes the size of an element within the mesh and $c$ denotes the speed of compressional waves in the medium under consideration. Further details on the numerical implementation are provided in SI. Further, videos of the growing crack front in the three benchmark cases are provided as multimedia supplementary information.\\

\begin{figure}[h]
	\centering
	\includegraphics[width=\textwidth]{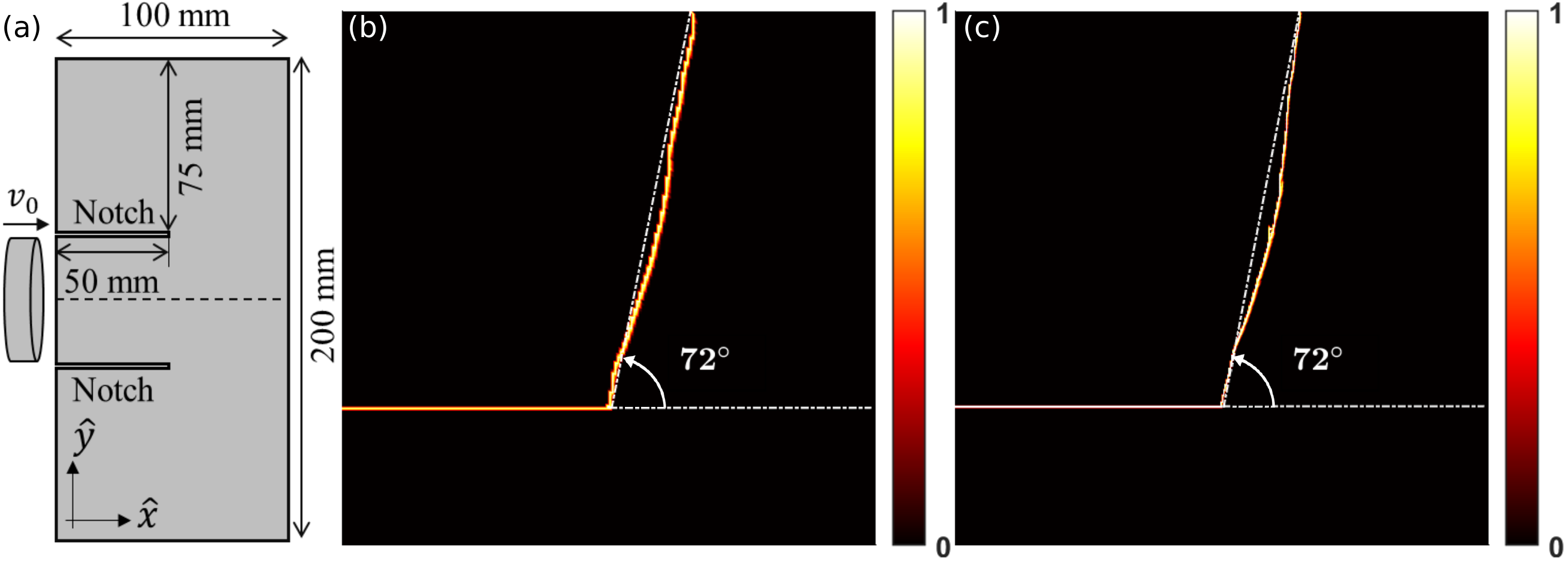}
	\caption{\label{fig: G3} (a) Schematic of the edge-cracked plate under impact loading known as the Kalthoff-Winkler experiment \cite{kalthoff1988failure}. Crack patterns simulated using the VO damage model are presented for two different mesh configurations (b) $h=0.5$ mm and (c) $h=0.25$ mm. The average angle of the crack front in both (b) and (c) is obtained as $72^\circ$ indicating an excellent match with the experimental result.}
\end{figure}

\noindent\textbf{Kalthoff-Winkler experiment:} The classical Kalthoff-Winkler experiment \cite{kalthoff1988failure} consists of an unrestrained doubly notched specimen subject to an impact load, as illustrated in Fig.~(\ref{fig: G3}a). Following the original experimental setup \cite{kalthoff1988failure}, the specimen was made of maraging steel with the following material properties: $E=190$ GPa, $\sigma_u=844$ MPa, $G_f=22.2$ N/mm, $\nu=0.3$, and $\rho=8000$ kg/m$^3$ \cite{cusatis2017discontinuous}. The characteristic material length corresponding to the aforementioned material properties is obtained as $l_t=0.012$ m. It was observed experimentally that, for lower strain rates ($v_0=16.5$ m/s), brittle failure occurs and the cracks nucleate from the edges of both notches at an angle of about $70^\circ$ with respect to the horizontal axis (which coincides with the line of symmetry). In the following, we numerically analyze this benchmark problem using the VO dynamic fracture model.

Given the symmetry of both the specimen and the test conditions in the original experiment, we modeled only the upper-half of the specimen in order to reduce the computational cost. The vertical component of the displacement field was set as zero ($u_y$=0) at the line of symmetry indicated in Fig.~(\ref{fig: G3}a) to impose the symmetric boundary conditions. To model the impact load, a velocity boundary condition was applied at the nodes corresponding to the impact zone and the impulse was kept constant throughout the dynamic simulation. Further, the linear softening law (see Eq.~(\ref{eq: softening_law_linear})) was used to model the degradation in the elastic energy upon damage development. The damage pattern generated using the VO model is presented in Fig.~(\ref{fig: G3}b,\ref{fig: G3}c) for two different mesh configurations. The element size is taken as $h=0.5$ mm in Fig.~(\ref{fig: G3}b), and $h=0.25$ mm in Fig.~(\ref{fig: G3}c). As evident from Figs.~(\ref{fig: G3}b,\ref{fig: G3}c), although a sharper crack is obtained in the case of the finer mesh, the overall crack propagation features are insensitive to the specific mesh configuration. For both mesh configurations, the crack develops in a direction that forms approximately $70^\circ$ with the horizontal axis. The average angle from the initial crack tip to the point where the crack intersects the top boundary is obtained as $72^\circ$ which matches well with the experimental result in \cite{kalthoff1988failure}. Further, the crack intersects the top boundary of the specimen at a time instant of 75 $\mu$s for both the mesh configurations; this time corresponds to an average crack propagation velocity of $\overline{c}=1064$ m/s. Note that $\overline{c}<0.6c_R$, where $c_R$ ($=2745$ m/s) denotes the Rayleigh wave speed in the medium, as observed commonly in experiments \cite{ravi1984experimental}. For completeness, we highlight that unlike the results presented in \cite{song2008comparative,cusatis2017discontinuous} and obtained using different FEM and DCM models, we did not observe any spurious cracks developing from the bottom right corner and travelling towards the tip of the notch.\\

\begin{figure}[h]
	\centering
	\includegraphics[width=\textwidth]{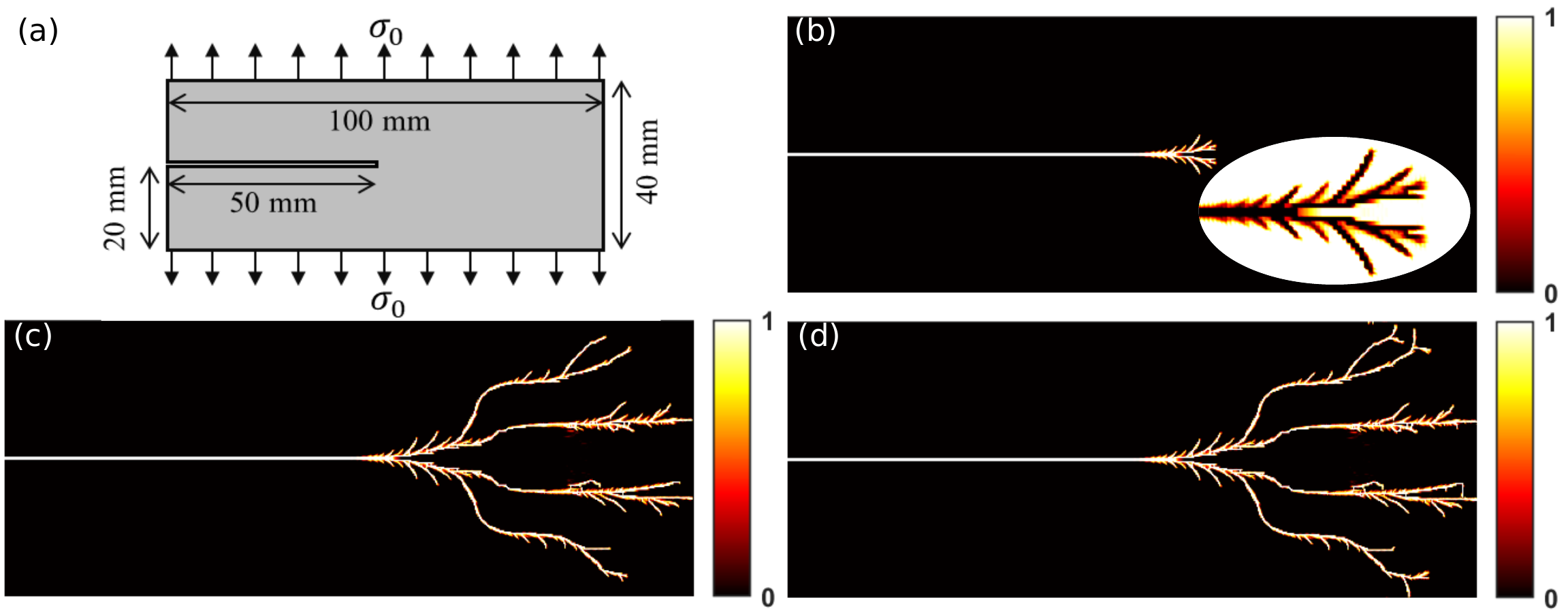}
	\caption{\label{fig: G2}(a) Schematic of the edge cracked block under tensile loading \cite{ravi1984experimental}. (b) Numerical simulation of the crack front near the onset of instability. The roughening of the crack surface is shown in the inset that provides a zoomed-in view of the crack face (color scheme inverted for better visibility). (c,d) Crack patterns obtained at the instants when the innermost branches and the outermost branches reach the boundaries of the specimen.}
\end{figure}

\noindent\textbf{Dynamic crack branching:} In this benchmark problem, we model a pre-cracked specimen loaded dynamically in tension as illustrated in Fig.~(\ref{fig: G2}a). This problem has been widely adopted in the literature to study dynamic crack branching, both experimentally \cite{ravi1984experimental} and numerically \cite{rabczuk2004cracking,song2008comparative,borden2012phase,cusatis2017discontinuous,nguyen2018modeling}. The specific material parameters used in this simulation were $E=32$ GPa, $\sigma_u = 3.1$ MPa, $G_f=3$ J/m$^2$, $\nu=0.2$ and $\rho=2450$ kg/m$^3$ \cite{cusatis2017discontinuous}, which correspond to a glass type material. The associated characteristic length is found to be $l_t=0.02$ m. Further, given the brittle nature of the material in this experiment, the linear softening law was used to account for the degradation in the elastic strain energy upon damage. A uniform and constant traction of magnitude $\sigma_0=1$ MPa is applied instantaneously to the rectangular specimen on its top and bottom surfaces at the initial time step and is held constant throughout the simulation. All other surfaces of the specimen are left free. This load condition is such that crack branching occurs in the specimen. The simulations obtained via the VO formulation are presented in Fig.~(\ref{fig: G2}b-\ref{fig: G2}d). Figure~(\ref{fig: G2}b) depicts the crack pattern at an instant following the onset of instability, and Fig.~(\ref{fig: G2}c) and Fig.~(\ref{fig: G2}d) present the crack pattern when the innermost and outermost branches reach the boundaries of the specimen, respectively.

The results obtained via the VO model lead to the following remarks. First, as discussed in \cite{ravi1984experimental}, upon the onset of instability several microcracks develop from the principal propagating crack branch and interact with one another simultaneously. This process ultimately leads to a roughening of the crack surface. As evident from the inset in Fig.~(\ref{fig: G2}b), the VO formulation is able to capture in great detail the roughening of the crack surface due to the emerging microbranches. Similar to the experiments conducted in \cite{ravi1984experimental}, these microbranches vary in size and the larger ones develop into full fledged branches. The remaining microbranches are arrested as a result of dynamic interaction with the growing ones. This set of results also highlights some of the advantages of the VO model over phase-field models which invariably capture an \textit{artificial} widening effect in the damaged area at the point of occurrence of instability \cite{borden2012phase,nguyen2018modeling}.
Also extremely remarkable, unlike classical dynamic fracture models that capture two branches \cite{song2008comparative,borden2012phase,cusatis2017discontinuous,nguyen2018modeling}, the VO model predicts four branches nucleating from the point of instability. This result closely matches the experimental results in \cite{ravi1984experimental}, where it was demonstrated that the number of branches developed can vary between two and four.
Further, we emphasize that unlike classical discrete approaches to dynamic fracture \cite{belytschko2003dynamic,remmers2003cohesive,song2008comparative,park2012adaptive} we did not impose any additional criteria within the VO model to facilitate the crack branching behavior; the branching and roughening occurs naturally as a result of the local response field. Similar to phase field (variational) approaches, the VO dynamic fracture model leads to full crack identification without the support of additional branching conditions. 
Finally, as shown in \cite{ravi1984experimental}, the number of branches exceeds four when also considering unsuccessful (i.e. not fully developed) branches. This feature is also captured by the VO model wherein we see that a number of unsuccessful branches nucleate from the principal branches.\\

\begin{figure}[h]
	\centering
	\includegraphics[width=\textwidth]{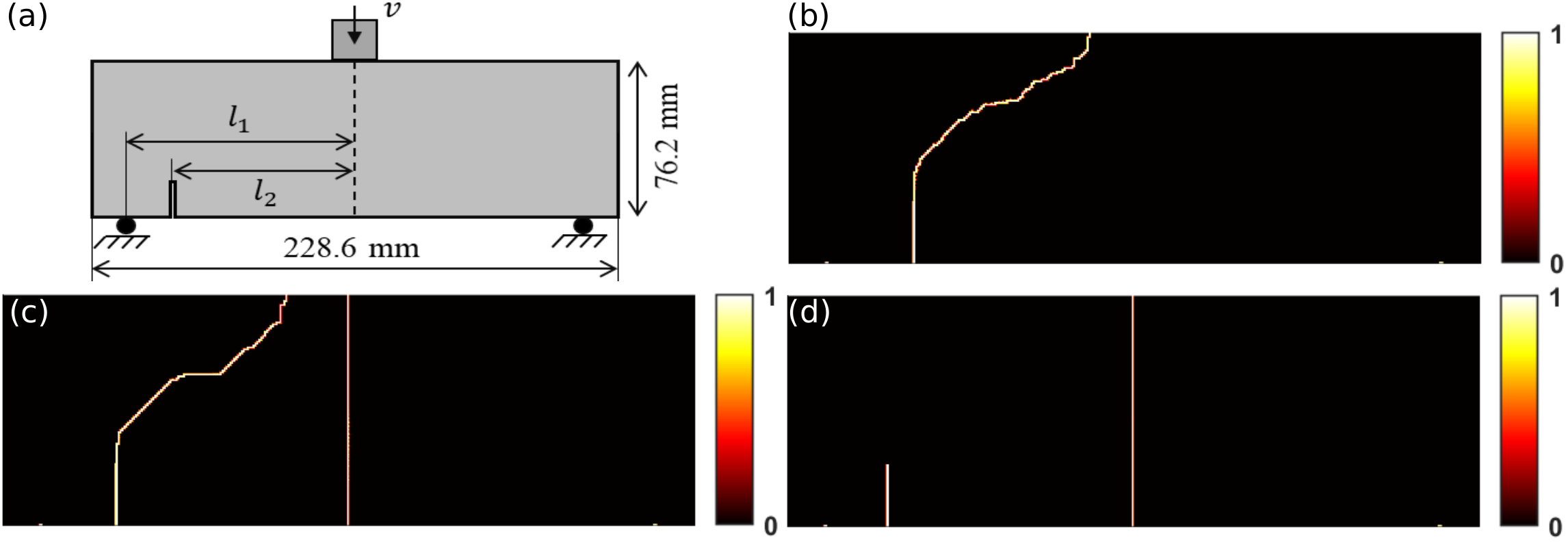}
	\caption{\label{fig: G4}(a) Schematic of the John-Shah experiment \cite{john1990mixed} involving the impact loading of a pre-notched concrete beam resting on simple supports. The length of the initial notch is $19.4$ mm. The numerically obtained crack patterns for different notch locations (b) $\gamma=0.72$, (c) $\gamma=0.76$, and (d) $\gamma=0.79$. Similar to the experiment, the crack nucleates from the tip of the notch for $\gamma=0.72<\gamma_c$, from the mid-span of the beam for $\gamma=0.79>\gamma_c$, and from both the notch tip and mid-span for $\gamma=0.76$.}
\end{figure}

\noindent\textbf{John-Shah experiment:} Another benchmark problem to test the performance of the VO fracture mechanics framework involves the three-point bending of concrete beams subject to impact loading \cite{john1990mixed}. The geometry and boundary conditions for the specimen involved in this test are illustrated in Fig.~(\ref{fig: G4}a). In this classical benchmark problem, a pre-built notch (offset from the mid-span axis of symmetry) is used to study mixed-mode fracture in concrete beams. It was observed by John and Shah that the parameter $\gamma=l_1/l_2$ (see Fig.~(\ref{fig: G4}a)), that controls the placement of the notch, plays a critical role in determining what failure mode and damage pattern would occur in the specimen after the impact. Indeed, there exits a critical value $\gamma_c$ such that, for $\gamma<\gamma_c$, the crack nucleates from the notch tip while, for $\gamma>\gamma_c$, the crack nucleates from the mid-span. In addition, there exists an intermediate value of $\gamma$ close to and less than $\gamma_c$ wherein both cracks develop. The experimentally determined value of $\gamma_c$ was $\gamma_c=0.77$ \cite{john1990mixed}.

We simulated this benchmark problem using the VO framework. The specific material parameters used in the simulation were $E=34$ GPa, $\sigma_u = 1$ MPa, $G_f=31.1$ J/m$^2$, $\nu=0.2$ and $\rho=2400$ kg/m$^3$. The degradation in the strain energy upon damage was modeled using the Cornelissen softening law for concrete (see Eq.~(\ref{eq: softening_law_cornelissen})). The impact velocity is given by the linear ramp \cite{belytschko2000element}:
\begin{equation}
    \label{eq: G4_VBC}
    v(t)=\left\{ \begin{matrix}  \frac{v_0 t}{t_0} & t\leq t_0 \\ v_0 & t>t_0\end{matrix}\right.
\end{equation}
where $v_0=0.06$ m/s and $t_0=196~\mu$s. 
Using the above material properties, geometry, and loading conditions we simulated the dynamic three-point bending for three different values of $\gamma\in\{0.72,0.76,0.79\}$ corresponding to three different notch locations. In all the three cases, the computational domain was uniformly discretized using elements of size $h=0.635$ mm. Note that the characteristic material length corresponding to the material properties of the concrete specimen is obtained as $l_t=2.1$~m. We merely observe that for quasi-brittle materials like concrete, the characteristic length-scale is generally too large when compared to the dimensions of laboratory specimens and hence, the condition $l_f<l_t$ is virtually meaningless for quasi-brittle materials.
The crack obtained for the three different cases are presented in Figs.~(\ref{fig: G4}b-\ref{fig: G4}d).

Overall, the results of the three numerical experiments compare very well with the experimental results. In particular, as in the experimental results, the crack nucleates from the tip of the notch for $\gamma=0.72<\gamma_c$ and from the mid-span of the beam for $\gamma=0.79>\gamma_c$, and propagates towards the top surface of the beam. Further, similar to the experiment conducted in \cite{john1990mixed}, a transition state is observed for the case $\gamma=0.76$ wherein cracks propagate from both the notch tip and mid-span towards the top surface. It follows that the estimate of the critical notch location $\gamma_c$ obtained via the VO dynamic fracture model lies in $(0.76,0.79)$, which is in good agreement with the experimental value of $\gamma_c=0.77$.

\section*{Conclusions}

This study presented a novel elastodynamic formulation based on variable order fractional operators and capable to provide accurate estimates of dynamic fracture in brittle and quasi-brittle solids. From a mathematical perspective, the peculiar properties of the VO-RL operator enable capturing the behavior of highly nonlinear systems with evolving discontinuities, such as those involved in the nucleation and propagation of cracks in solids. We showed that an apparently unsettling property of the RL operator, that is the non-vanishing derivative of a constant, can have very useful implications to model dynamic fracture. Furthermore, the ability of VO operators to update their order as a function of either dependent or independent variables, results in governing equations that can evolve in real time to capture growing cracks without requiring modifications to the fundamental governing equations. Even more remarkable, and certainly in stark contrast with more traditional approaches to dynamic fracture, is the fact that VO governing equations do not require \textit{a priori} assumptions or additional conditions to detect characteristic aspects of dynamic fracture such as crack nucleation, crack surface roughening, crack instability and branching. In other terms, the nonlinear and discontinuous dynamic behavior associated with fracture naturally emerges based on the instantaneous response of the system. Further, given the many recent advances in the formulation of fractional order mechanics as a comprehensive approach to nonlocal elasticity, it can be envisioned that the present VO elastodynamic framework could be easily integrated in a fully fractional formulation hence leading to a powerful tool for dynamic fracture analysis of nonlocal media.\\


\noindent\textbf{Acknowledgements:} The authors gratefully acknowledge the financial support of the Defense Advanced Research Project Agency (DARPA) under grant \#D19AP00052, and of the National Science Foundation (NSF) under grants MOMS \#1761423 and DCSD \#1825837. The content and information presented in this manuscript do not necessarily reflect the position or the policy of the government. The material is approved for public release; distribution is unlimited.\\

\noindent\textbf{Competing Interests:} The authors declare that there are no competing interests.\\

\noindent\textbf{Author Contribution:} All authors have contributed equally to the work.\\

\noindent\textbf{Data Availability:} The algorithm is entirely described in SI with all the details necessary to reproduce the model. The files including the numerical results as well as videos corresponding to the results in the paper will also be shared. The actual physical code might be subject to restrictions from the sponsor and, at this stage, we cannot guarantee we will be allowed to share it.\\

\bibliographystyle{unsrt}
\bibliography{report}

\end{document}